\documentstyle[12pt,graphicx,epsf,braket]{article}
\begin{document}

\def\AEF{A.E. Faraggi}

\def\vol#1#2#3{{\bf {#1}} ({#2}) {#3}}
\def\NPB#1#2#3{{\it Nucl.\ Phys.}\/ {\bf B#1} (#2) #3}
\def\PLB#1#2#3{{\it Phys.\ Lett.}\/ {\bf B#1} (#2) #3}
\def\PRD#1#2#3{{\it Phys.\ Rev.}\/ {\bf D#1} (#2) #3}
\def\PRL#1#2#3{{\it Phys.\ Rev.\ Lett.}\/ {\bf #1} (#2) #3}
\def\PRT#1#2#3{{\it Phys.\ Rep.}\/ {\bf#1} (#2) #3}
\def\MODA#1#2#3{{\it Mod.\ Phys.\ Lett.}\/ {\bf A#1} (#2) #3}
\def\RMP#1#2#3{{\it Rev.\ Mod.\ Phys.}\/ {\bf #1} (#2) #3}
\def\IJMP#1#2#3{{\it Int.\ J.\ Mod.\ Phys.}\/ {\bf A#1} (#2) #3}
\def\nuvc#1#2#3{{\it Nuovo Cimento}\/ {\bf #1A} (#2) #3}
\def\RPP#1#2#3{{\it Rept.\ Prog.\ Phys.}\/ {\bf #1} (#2) #3}
\def\APJ#1#2#3{{\it Astrophys.\ J.}\/ {\bf #1} (#2) #3}
\def\APP#1#2#3{{\it Astropart.\ Phys.}\/ {\bf #1} (#2) #3}
\def\EJP#1#2#3{{\it Eur.\ Phys.\ Jour.}\/ {\bf C#1} (#2) #3}
\def\etal{{\it et al\/}}

\newcommand{\cc}[2]{c{#1\atopwithdelims[]#2}}
\newcommand{\bev}{\begin{verbatim}}
\newcommand{\beq}{\begin{equation}}
\newcommand{\beqa}{\begin{eqnarray}}
\newcommand{\beqn}{\begin{eqnarray}}
\newcommand{\eeqn}{\end{eqnarray}}
\newcommand{\eeqa}{\end{eqnarray}}
\newcommand{\eeq}{\end{equation}}
\newcommand{\beqt}{\begin{equation*}}
\newcommand{\eeqt}{\end{equation*}}
\newcommand{\Eev}{\end{verbatim}}
\newcommand{\bec}{\begin{center}}
\newcommand{\eec}{\end{center}}
\def\ie{{\it i.e.}}
\def\eg{{\it e.g.}}
\def\half{{\textstyle{1\over 2}}}
\def\nicefrac#1#2{\hbox{${#1\over #2}$}}
\def\third{{\textstyle {1\over3}}}
\def\quarter{{\textstyle {1\over4}}}
\def\m{{\tt -}}
\def\mass{M_{l^+ l^-}}
\def\p{{\tt +}}

\def\slash#1{#1\hskip-6pt/\hskip6pt}
\def\slk{\slash{k}}
\def\GeV{\,{\rm GeV}}
\def\TeV{\,{\rm TeV}}
\def\y{\,{\rm y}}

\def\l{\langle}
\def\r{\rangle}
\def\LRS{LRS  }

\begin{titlepage}
\samepage{
\setcounter{page}{1}
\rightline{LTH--915} 
\rightline{arXiv:????.????}
\vspace{1.5cm}
\begin{center}
 {\Large \bf Proton Stability \\ and \\
\medskip Light $Z^\prime$ Inspired by
             String Derived Models}
\vspace{.25 cm}

Alon E. Faraggi\footnote{
		                  E-mail address: faraggi@amtp.liv.ac.uk}
and 
Viraf M. Mehta\footnote{ E-mail address:
	                          Viraf.Mehta@liv.ac.uk}
\\
\vspace{.25cm}
{\it Department of Mathematical Sciences\\
University of Liverpool, Liverpool, L69 7ZL, United Kingdom}
\end{center}

\begin{abstract}

Proton stability is one of the most perplexing puzzles in 
particle physics. While the renormalizable Standard Model forbids
proton decay mediating operators due to accidental global symmetries,
many of its extensions introduce such dimension four, five and six operators. 
Furthermore, it is, in general, expected that quantum gravity only respects
local gauge, or discreet, symmetries. String theory provides the 
arena to study particle physics in a consistent framework of perturbative 
quantum gravity. An appealing proposition, in this context, is that 
the dangerous operators are suppressed by an Abelian gauge symmetry, which is
broken near the TeV scale. A viable $U(1)$ symmetry should also be
anomaly free, be family universal,
and allow the generation of fermion masses via
the Higgs mechanism. 
We discuss such $U(1)$ symmetries that arise in 
quasi--realistic free fermionic heterotic--string derived models.
Ensuring that the $U(1)$ symmetry is anomaly free at the low scale 
requires that the Standard Model spectrum is augmented by
additional states that are compatible with the charge
assignments in the string models. We construct such 
string--inspired models and discuss some of their 
phenomenological implications.

\end{abstract}
\smallskip}
\end{titlepage}

\section{Introduction}

~~Proton longevity is one of the important clues in attempts to understand
the fundamental origins of the basic constituents of matter and 
their interactions. In the Standard Model baryon and lepton numbers are
protected at the renormalizable level by accidental global symmetries.
However, the Standard Model is merely an effective field theory 
below some cutoff scale. Nonrenormalizable operators suppressed 
by the cutoff scale will, in general, violate baryon and lepton numbers,
unless they are forbidden by additional symmetries.
Indeed in the Standard Model, such
proton decay mediating operators appear at dimension six
and indicate that the cutoff scale is above $10^{16}$GeV \cite{psreview}. 

Many extensions of the Standard Model that
have been proposed to address other issues, in particular the hierarchy
problem, introduce a cutoff at the TeV scale. Such extensions 
consequently induce proton decay at an unacceptable rate.
One must then rely on some ad hoc global or discrete symmetries,
that forbid the unwanted terms. In general, it is expected that only
local symmetries are not violated by quantum gravity
effects \cite{qge}. Therefore, an appealing proposition is that the
suppression of the proton decay mediating operators 
is due to the existence of an Abelian gauge symmetry beyond the
Standard Model. For the extra symmetry to provide adequate
suppression of the unwanted terms, it has to exist at a mass
scale within reach of contemporary particle accelerators \cite{ps}. 

In supersymmetric extensions of the Standard Model baryon 
and lepton number violating dimension four operators are
given by$$QLD~{\rm and} ~UDD,$$ 
where $Q$ is the quark left--handed electroweak doublet; $L$ 
is the lepton left--handed electroweak doublet; and $U$ and $D$ 
are the anti--quark, up and down, left--handed electroweak 
singlets \cite{wei}. 
These operators are forbidden by gauged baryon minus lepton 
number, $U(1)_{B-L}$, which arises in 
Pati--Salam \cite{patisalam} and $SO(10)$
Grand Unified Theories. $SO(10)$ grand unification is an appealing 
extension of the Standard Model, as each of the Standard Model
generations is embedded in a single $SO(10)$ spinorial representation.
The requirement of light neutrino masses, however, necessitates that 
lepton number is broken. In $SO(10)$ grand unified models one can 
use the 126 representation, which breaks lepton number by two 
units and leaves an unbroken symmetry, which still forbids the 
dimension four operators. However, the 126 representation,
in general \cite{dmr} does not arise in perturbative string models. 
This implies that lepton number is broken by unit one
carrying fields that generate the dangerous dimension four 
operators. Specifically, in $SO(10)$ these operators will arise 
from the $16^4$ term, that gives rise to the dimension five terms
$$QLDN~{\rm and}~ UUDN,$$
where $N$ is the Standard Model singlet field and gets a VEV of 
the order of the GUT scale. Additionally, the $16^4$ 
gives rise to the dimension five terms 
$$QQQL ~{\rm and }~ UUDE,$$
where $E$ is the charged anti--lepton left--handed electroweak singlet.
These dimension five operators are not forbidden by $U(1)_{B-L}$.
It is therefore apparent that gauged baryon minus lepton number 
by itself is not sufficient to guarantee proton stability. 
Other local gauge symmetries, possibly in conjunction with $U(1)_{B-L}$,
are needed to ensure proton stability \cite{custodial,pati}.

The existence of extra local Abelian symmetries beyond the Standard Model,
in GUTs and string theories, have been amply discussed in the literature
\cite{zprimeliterature}. Most appealing in this respect are the Abelian
extension that arise in $SO(10)$ and $E_6$ \cite{so10e6zprime}.
The embedding of the Standard Model matter 
in spinorial 16 $SO(10)$ representations, strongly hints that
the $SO(10)$ group structure is realised in nature, 
whereas $E_6$ goes a step further by accommodating the 
matter and Higgs states in common representations. 
Heterotic--string models produce gauge and matter
states that can be identified with the Standard Model particles
and admit their embedding in $SO(10)$ and $E_6$ multiplets.
A class of three generation heterotic--string models that produces
the Grand Unification embedding are the free fermion 
models \cite{fsu5,fny,alr,slm,lrs, exophobic},
and correspond to compactifications on 
$Z_2\times Z_2$ orbifolds \cite{xmap}. 

To date, the discussion of $U(1)$ symmetries as proton lifeguards in
free fermion string models has focused on the existence of such symmetries
\cite{lepzprime,pati,ps}, and
the condition that they need to satisfy
to remain viable at the string scale,
as well as some constraints emanating from the 
Standard Model data \cite{ps}. These include the requirements that the 
extra $U(1)$: forbids the dimension four, five and six proton decay
mediating operators; allows suppression of left--handed 
neutrino masses by a seesaw mechanism; allows the fermion
Yukawa couplings to electroweak Higgs doublets; 
is family universal; is anomaly free.
However, satisfying these requirements at the string scale does not 
guarantee that they are satisfied at the electroweak scale,
in particular, with respect to freedom from anomalies. 
The existence of the desired symmetry in explicit string constructions 
guarantees that in these examples the $U(1)$ symmetry is free of any 
gauge and gravitational anomalies. However, to facilitate the 
analysis of the phenomenological properties of the extra $U(1)$, 
what is needed is a toy string--inspired, field theory model
that takes into account the ingredients, in particular with respect
to the charges, from the string derived models. 

In this paper we undertake the task of constructing such string--inspired
models. The extra $U(1)$s that we consider satisfy the requirements 
listed above, while taking into account the charges as they arise in the 
string models. The spectrum in the string--inspired model is
constructed to satisfy anomaly cancellation at the electroweak scale.
We outline the conditions, as seen in the free fermionic models,
that are needed in order for the extra $U(1)$ to remain viable at low
energies. In terms of the charges of the matter states under the extra 
$U(1)$, there are broadly two classes of models: Those in which the 
charges satisfy an $E_6$ embedding; and those in which they do not. 
We elaborate on the two types of models and how they arise in the string
constructions. 


\section{The structure of the free fermionic models}\label{review}

In this section we review the construction and structure of the free fermionic
models. We particularly focus on the general gauge structure and the 
charges under the $U(1)$ symmetries that are identified with the 
low scale $Z^\prime$. A recurring feature of the free fermionic models
is the existence of an anomalous $U(1)$,
which often coalesce with the
$U(1)$ symmetry that serves as the proton lifeguard \cite{cleaverau1}.
An important issue
on which we elaborate, is the conditions under which this $U(1)$
symmetry is anomaly free. 
Further details of the free fermionic models
are given in refs. \cite{fff,fny,slm,nahe}.

In the free fermionic formulation of the heterotic string
in four dimensions \cite{fff}, all the world--sheet
degrees of freedom,  required to cancel
the conformal anomaly, are represented in terms of free fermions
propagating on the string world--sheet.
In the light--cone gauge, the world--sheet free fermion fields 
consist
of two transverse left--moving space--time fermionic coordinates,
$\psi^\mu_{1,2}$, and an additional 62 purely internal
Majorana--Weyl fermions, of which 18 are left--moving,
and 44 are right--moving.
The models are constructed by specifying the phases picked by
the world--sheet fermions when transported along the torus
non--contractible loops
\beq  f \rightarrow -e^{i\pi\alpha(f)}f, \quad \alpha(f) \in (-1,1]. \eeq
Each model corresponds to a particular choice
of fermion phases consistent with modular invariance and is generated
by a set of basis vectors describing the transformation properties
of the 64 world--sheet fermions.
The boundary
conditions basis vectors ${b}_k$ span a finite additive group
\begin{equation}
\Xi={\sum_k}n_i {b}_i
\label{additivegroup}
\end{equation}
where $n_i=0,\cdots,{{N_{z_i}}-1}$.
The physical massless states in the Hilbert space of a given sector
$\alpha\in{\Xi}$ are then obtained by acting on the vacuum state of
that sector with the world-sheet bosonic and fermionic mode operators,
with frequencies $\nu_f$, $\nu_{f^*}$ and
by subsequently applying the Generalized GSO projections,
\begin{equation}
\left\{e^{i\pi({b_i}F_\alpha)}-
{\delta_\alpha}c^*\left(\matrix{\alpha\cr
                 b_i\cr}\right)\right\}\vert{s}\rangle=0
\label{gsoprojections}
\end{equation}
with
\begin{equation}
(b_i{F_\alpha})\equiv\{\sum_{real+complex\atop{left}}-
\sum_{real+complex\atop{right}}\}(b_i(f)F_\alpha(f)),
\label{lorentzproduct}
\end{equation}
where $F_\alpha(f)$ is a fermion number operator counting each mode of
$f$ once (and if $f$ is complex, $f^*$ minus once). For periodic
complex fermions [{\it i.e.} for $\alpha(f)=1)$]
the vacuum is a spinor in order to represent
the Clifford algebra of the corresponding zero modes.
For each periodic complex fermion $f$,
there are two degenerate vacua $\vert{+}\rangle$, $\vert{-}\rangle$,
annihilated by the zero modes $f_0$ and $f^*_0$ and with fermion
number $F(f)=0,-1$ respectively. 
In Eq. (\ref{gsoprojections}),
$\delta_\alpha=-1$ if $\psi^\mu$ is periodic in the sector $\alpha$,
and $\delta_\alpha=+1$ if $\psi^\mu$ is antiperiodic in the sector 
$\alpha$.
The $U(1)$ charges with respect to the unbroken Cartan generators of the
four dimensional gauge group are in one to one correspondence with the
$U(1)$ $ff^*$ currents. For each complex fermion $f$:
\begin{equation}
Q(f)={1\over2}\alpha(f)+F(f).
\label{u1numbers}
\end{equation}
The representation (\ref{u1numbers})
shows that $Q(f)$ is identical with the world--sheet
fermion numbers $F(f)$ for world--sheet fermions with Neveu--Schwarz
boundary conditions $(\alpha(f)=0)$,
and is $F(f)+{1\over2}$ for those with Ramond boundary conditions
$(\alpha(f)=1)$.
The charges for the $\vert\pm\rangle$ spinor vacua are $\pm{1\over2}$.

The sixty--four world--sheet fermions in the light--cone gauge
are divided in the following way:\begin{itemize}

\item   a complex right-moving fermion, denoted $\psi^\mu$, 
         formed from the two
         real fermionic superpartners of the coordinate
         boson $X^\mu$;

\item   six real right-moving fermions denoted $\chi^{1,...,6}$,
         often paired to form three complex right-moving fermions
         denoted $\chi^{12}$, $\chi^{34}$, and $\chi^{56}$;

\item   12 real right-moving fermions, denoted $y^{1,...,6}$
          and $\omega^{1,...,6}$;

\item   12 corresponding real left-moving fermions, denoted
$\overline{y}^{1,...,6}$
          and $\overline{\omega}^{1,...,6}$; and

\item   16 remaining complex left-moving fermions, denoted
          $\overline{\psi}^{1,...,5}$,  $\overline{\eta}^{1,...,3}$,
          and $\overline{\phi}^{1,...,8}$.

\end{itemize}
The complex right--moving fermions
${\bar\phi}^{1,\cdots,8}$ generate the rank eight hidden gauge group; 
${\bar\psi}^{1,\cdots,5}$ generate the $SO(10)$ GUT gauge group;
${\bar\eta}^{1,2,3}$ generate the three remaining $U(1)$
generators in the Cartan sub-algebra of
the observable rank eight gauge group.
A combination of these three $U(1)$ currents plays the role of
the proton lifetime guard \cite{ps}. 

Each free fermion model is defined in terms
of a set of basis vectors ${b}_i,i=1,\dots,n$,
and the one--loop GGSO projection coefficients $\cc{b_i}{b_j}$.
There are two broad classes of free fermionic models that 
have been studied. The first class are models
that utilise the NAHE set of boundary condition basis vectors. 
The second class are the models spanned in the classification 
of refs. \cite{fkr}. The two classes differ in that the first
allows and uses complexified internal fermions from the set
$\{y, \omega\vert {\bar y},{\bar\omega}\}$, whereas 
such fermions have not been incorporated in the
second class to date. Complexification of internal fermions results in 
additional $U(1)$ gauge symmetries in the first class of models. 
The treatment of the sixteen complex world--sheet fermions 
that generate the 
gauge degrees of freedom is identical in the two classes of models. 
As the extra proton safeguarding $U(1)$ symmetry
arises exclusively from these world--sheet fermions,
the two classes are identical in respect to the
extra $Z^\prime$ of interest here.
To date, the majority of phenomenological studies of free fermionic models
are NAHE based \cite{fsu5,fny,alr,slm,lrs,su421}, with the 
notable exception being the exophobic Pati--Salam vacua of refs.
\cite{exophobic}. 
For definiteness, we discuss, in this paper, the NAHE--based models.
The first stage in the construction of these models 
consists of the NAHE--set of basis vectors,
$\{{\bf 1}, S,b_1,b_2,b_3\}$ \cite{nahe}.
The gauge group at this stage is $SO(10)\times SO(6)^3\times E_8$,
and the vacuum contains forty--eight multiplets in the 16 chiral
representation of $SO(10)$. The second stage consists of adding
to the NAHE--set three or four basis vectors, typically denoted by
$\{\alpha,\beta,\gamma\}$. The additional basis vectors
reduce the number of generations to three, 
and break the four dimensional gauge symmetry. 
The $SO(10)$ GUT group is broken to one of its subgroups
by the following assignment of boundary conditions to the set of
complex world--sheet fermions ${\bar\psi}^{1,\cdots,5}_{1\over2}$:

\beqn
&1.&b\{{{\bar\psi}_{1\over2}^{1\cdots5}}\}=
\{{1\over2}{1\over2}{1\over2}{1\over2}
        {1\over2}\}\Rightarrow SU(5)\times U(1),\label{su51breakingbc}\\
&2.&b\{{{\bar\psi}_{1\over2}^{1\cdots5}}\}=\{1~ 1\, 1\, 0\, 0\}\,
  \Rightarrow SO(6)\times SO(4).
\label{su51so64breakingbc}
\eeqn
To break the $SO(10)$ symmetry to
$SU(3)_C\times SU(2)_L\times
U(1)_C\times U(1)_L$ \cite{slm}
both steps, 1 and 2, are used, in two separate basis
vectors\footnote{$U(1)_C={3\over2}U(1)_{B-L};
U(1)_L=2U(1)_{T_{3_R}}.$}.
The breaking pattern
$SO(10)\rightarrow SU(3)_C\times SU(2)_L\times SU(2)_R \times U(1)_{B-L}$
\cite{lrs} is achieved by the following assignment in two separate basis
vectors
\beqn
&1.&b\{{{\bar\psi}_{1\over2}^{1\cdots5}}\}=\{1~1\, 1\, 0 0\}
  \Rightarrow SO(6)\times SO(4),\\
&2.&b\{{{\bar\psi}_{1\over2}^{1\cdots5}}\}=
\{{1\over2}{1\over2}{1\over2}00\}\Rightarrow SU(3)_C\times U(1)_C
\times SU(2)_L\times SU(2)_R.
\label{su3122breakingbc}
\eeqn
Similarly, the breaking pattern
$SO(10)\rightarrow SU(4)_C\times SU(2)_L\times U(1)_R$ \cite{su421}
is achieved by the following assignment in two separate basis
vectors

\beqn
&1.& b\{{{\bar\psi}_{1\over2}^{1\cdots5}}\}=\{1 1 1\, 0 0\}
  \Rightarrow SO(6)\times SO(4),\\
&2.& b\{{{\bar\psi}_{1\over2}^{1\cdots5}}\}=
\{000{1\over2}{1\over2}\}\Rightarrow SU(4)_C\times
SU(2)_L\times U(1)_R.
\label{su421breakingbc}
\eeqn
It was demonstrated in \cite{su421} that the breaking pattern
(\ref{su421breakingbc}) does not produce viable models in NAHE--based
models. 
Viable three generation models with: 
$SU(5)\times U(1)$ \cite{fsu5};
$SO(6)\times SO(4)$ \cite{alr}; 
$SU(3)\times SU(2)\times U(1)^2$ \cite{slm}; or
$SU(3)\times SU(2)^2\times U(1)$ \cite{lrs}, 
$SO(10)$ sub--groups were constructed.  
Three chiral generations arise
from the sectors $b_1$, $b_2$ and $b_3$ and are decomposed under the
final $SO(10)$ subgroup. 
The flavour $SO(6)^3$ groups are broken to products of $U(1)^n$ with
$3\le n\le 9$. The $U(1)_{1,2,3}$ factors arise from the three 
right--moving complex fermions ${\bar\eta}^{1,2,3}$. 

In order to elucidate the $U(1)_{1,2,3}$ charges of 
the matter states
in the free fermionic models, it is instructive to extend the $SO(10)$ symmetry,
at the level of the NAHE set, to $E_6$. This is achieved by adding to the
NAHE set the basis vector \cite{xmap}
\beq
x\equiv~ \{{\bar\psi}^{1,\cdots,5}{\bar\eta}^{1,2,3}\}\equiv 1.
\label{vectorx}
\eeq
With an appropriate choice of the GGSO projection coefficients,
the $x$--sector produces space--time vector bosons
that transform as $16\oplus\overline{16}$ under $SO(10)$ and
extend the $SO(10)\times U(1)$ symmetry to $E_6$.  The $U(1)$ 
combination is given by 
\beq
U(1)_\zeta=U_1+U_2+U_3
\label{u1zeta}
\eeq
where $U(1)_1$, $U(1)_2$ and $U(1)_3$
are the $U(1)$ symmetries generated by 
the right--moving complex world--sheet fermions 
${\bar\eta}^1$, 
${\bar\eta}^2$ and
${\bar\eta}^3$. 

In this model the gauge group is $SO(4)^3\times
E_6\times U(1)^2\times E_8$ with $N=1$ space--time supersymmetry.
There are 24 generations in the 27 representation of
$E_6$, eight from each twisted sector. In the fermionic construction
these are the sectors $(b_1;b_1+x)$, $(b_2;b_2+x)$ and $(b_3;b_3+x)$,
where the sectors $b_j$ produce the spinorial 
16 of $SO(10)$ and the sectors $b_j+x$ produce
the vectorial $(10 \oplus1) +1$ representations 
in the decomposition of the $27$ representation of $E_6$
\beq
27= 16_{1\over2} + 10_{-1} + 1_{2}
\label{27decom}
\eeq
under $SO(10)\times U(1)$. 
The additional ``1'' arising in the $b_j+x$ sectors
is an $E_6$ singlet. 

The vacuum of the sectors
$b_j$  contains twelve periodic fermions. Each periodic fermion
gives rise to a two dimensional degenerate vacuum $\vert{+}\rangle$ and
$\vert{-}\rangle$ with fermion numbers $0$ and $-1$, respectively.
After applying the
GSO projections, we can write the degenerate vacuum of the sector
$b_1$ in combinatorial form
\begin{eqnarray}
{\left[\left(\matrix{4\cr
                                    0\cr}\right)+
\left(\matrix{4\cr
                                    2\cr}\right)+
\left(\matrix{4\cr
                                    4\cr}\right)\right]
\left\{\left(\matrix{2\cr
                                    0\cr}\right)\right.}
&{\left[\left(\matrix{5\cr
                                    0\cr}\right)+
\left(\matrix{5\cr
                                    2\cr}\right)+
\left(\matrix{5\cr
                                    4\cr}\right)\right]
\left(\matrix{1\cr
                                    0\cr}\right)}\nonumber\\
{+\left(\matrix{2\cr
                                    2\cr}\right)}
&{~\left[\left(\matrix{5\cr
                                    1\cr}\right)+
\left(\matrix{5\cr
                                    3\cr}\right)+
\left(\matrix{5\cr
                                    5\cr}\right)\right]\left.
\left(\matrix{1\cr
                                    1\cr}\right)\right\}}
\label{spinor}
\end{eqnarray}
where
$4=\{y^3y^4,y^5y^6,{\bar y}^3{\bar y}^4,
{\bar y}^5{\bar y}^6\}$, $2=\{\psi^\mu,\chi^{12}\}$,
$5=\{{\bar\psi}^{1,\cdots,5}\}$ and $1=\{{\bar\eta}^1\}$.
The combinatorial factor counts the number of $\vert{-}\rangle$ in the
degenerate vacuum of a given state.
The first term in square brackets counts the degeneracy of the
multiplets, being eight in this case. 
The two terms in the curly brackets correspond to the two
CPT conjugated
components of a Weyl spinor.  The first
term among those corresponds to the 16 spinorial representation
of $SO(10)$, and fixes the space--time chirality properties of the
representation, whereas the second corresponds to the CPT conjugated
anti--spinorial $\overline{16}$ representation.
The charge under the $U(1)$ symmetry generated by ${\bar\eta}_1$
is determined by its vacuum state, being a Ramond state in the
$\vert +\rangle$ vacuum for the degenerate vacuum in Eq. (\ref{spinor}).
Hence, in this case the $U(1)_1$ charge is $+{1\over2}$. 
Similar vacuum structure is obtained 
for the sectors $b_2$ and $b_3$ with 
$\{\chi_{3,4},y^{1,2},\omega^{5,6} \vert
   {\bar y}^{1,2},{\bar\omega}^{5,6}, {\bar\eta}^2\}$
and 
$\{ \chi_{5,6},\omega^{1,\cdots,4}\vert
    {\bar\omega}^{1,\cdots,4}, {\bar\eta}^3 \}$
respectively. 

The $10+1$ in the $27$ of $E_6$ are obtained from the sector $b_j+x$.
The effect of adding the vector $x$ to the sectors $b_j$
is to replace the periodic boundary conditions for
${\bar\psi}^{1,\cdots,5},{\bar\eta}^j$ with periodic boundary conditions
for ${\bar\eta}^{i,k}$ with $i\ne k\ne j$ and $i,j,k\in\{1,2,3\}$.
Consequently, massless states from the sectors $b_j+x$ are obtained
by acting on the vacuum with a fermionic oscillator. In the case that the
$SO(10)$ symmetry is enhanced to $E_6$ these are in the $(10+1)+1$
representations of $SO(10)$, where the first $SO(10)$ singlet is in
the $27$ representation of $E_6$, whereas the second
is a singlet of $E_6$. These are obtained by acting on the 
vacuum with the oscillators of the complex world--sheet fermions
$\{{\bar\psi}^{1,\cdots,5}{\bar\eta}^j\}$, which have Neveu--Schwarz
boundary conditions in the sectors $b_j+x$. 
If the space--time vector bosons that enhance the $SO(10)\times
U(1)$ symmetry to $E_6$ are projected out, either the spinorial $16$ or
the vectorial $(10+1)+1$, survive the GSO projections
at a given fixed point. By breaking the degeneracy
with respect to the internal fermions 
$\{ y, \omega\vert{\bar y},{\bar\omega}\}$
we can obtain spinorial and vectorial representations from the 
twisted sectors at different fixed points. A classification of
symmetric free fermionic heterotic string models along these lines
was done in refs. \cite{fkr}.  

When the $SO(10)\times U(1)$ symmetry is enhanced to $E_6$, 
the charges of the 
spinorial $16$, the vectorial $10$ and the singlet $1$, under the 
$U(1)_\zeta$ in eq. (\ref{u1zeta}), are fixed by the 
$E_6$ symmetry, as shown in eq. (\ref{27decom}). 
When the $E_6$ symmetry is
broken by the GGSO projections the $U(1)_\zeta$ 
charges are not restricted by the $E_6$ embedding,
and can take either sign. 
The $U(1)$ symmetry that serves as the proton lifetime guard
is a combination of the three $U(1)$ symmetries generated by the 
world--sheet complex fermions ${\bar\eta}^{1,2,3}$. The states from 
each of the sectors $b_1$, $b_2$ and $b_3$ are charged with respect
to one of these $U(1)$ symmetries, {\it i.e.} with respect to 
$U(1)_1$, $U(1)_2$ and $U(1)_3$, respectively. 
Consequently, the $U(1)$ combination in eq. (\ref{u1zeta})
is family universal. 

In the string derived models of ref.
\cite{fsu5,fny,alr,slm}, the $U(1)_{1,2,3}$ are anomalous.
Therefore, $U(1)_\zeta\equiv U(1)_A$ is also anomalous and must be
broken near the string scale. In the string derived left--right
symmetric models of ref \cite{lrs}, $U(1)_{1,2,3}$ are anomaly free
and hence the combination $U(1)_\zeta$ is also anomaly free.
It is this property of these models which allows this 
$U(1)$ combination to remain unbroken. 

It is instructive to study  
the characteristics of $U(1)_\zeta$ in the 
left--right symmetric string derived models \cite{lrs}, versus those
of $U(1)_A$ in the string derived models of refs. \cite{fsu5,fny,alr,slm}.
We note that both $U(1)_\zeta$ as well as $U(1)_A$
are obtained from the same combination of complex right--moving 
world--sheet currents $\bar\eta^{1,2,3}$, {\it i.e.} both are 
given by a combination of $U_1$, $U_2$, and $U_3$.
The distinction between the two cases, as we describe in detail below,
is due to the charges of
the Standard Model states, arising from the sectors $b_1$, $b_2$ and $b_3$,
under this combination.
The key feature of $U(1)_\zeta$ in the models
of ref. \cite{lrs} is that it is anomaly free.

The periodic boundary conditions of the world--sheet
fermions ${\bar\eta}^j$ ensures that the fermions from each sector
$b_j$ are charged with respect to one of the $U(1)_j$ symmetries.
The charges, however, depend on the $SO(10)$ symmetry breaking 
pattern, induced by the basis vectors that extend the NAHE--set, and
may, or may not, differ in sign between different components of a given
generation. In the models of ref. \cite{fsu5,fny, slm, alr} 
the charges of a given $b_j$ generation under $U(1)_j$ is of the 
same sign, whereas in the models of ref. \cite{lrs} they differ.
In general, the distinction is by the breaking of $SO(10)$ to
either $SU(5)\times U(1)$ or $SO(6)\times SO(4)$. In the former case
they will always have the same sign, whereas in the latter they may
differ. To see why, it is instructive to consider the decomposition 
of the spinorial $16$ representation under the Pati--Salam subgroup.

Using the combinatorial notation introduced in eq. (\ref{spinor})
the decomposition of the 16 representation of $SO(10)$ in the 
Pati--Salam string models is 
\beq
\{\left[\left(\matrix{3\cr
                                    0\cr}\right)+
\left(\matrix{3\cr
                                    2\cr}\right)\right]
\left[\left(\matrix{2\cr
                                    0\cr}\right)+
\left(\matrix{2\cr
                                    2\cr}\right)\right]\}+
\{\left[\left(\matrix{3\cr
                                    1\cr}\right)+
\left(\matrix{3\cr
                                    3\cr}\right)\right]
\left[\left(\matrix{2\cr
                                    1\cr}\right)\right]\}
\label{psspinor}
\eeq
The crucial point is that the Pati--Salam breaking pattern
allows the first and second terms in curly brackets to 
come with opposite charges under $U(1)_j$. This results from the 
operation of the GSO projection operator, Eq. (\ref{gsoprojections}),
which differentiates between the two terms. The Pati--Salam breaking
pattern is common to the Pati--Salam models \cite{alr}, 
the Standard--like models \cite{fny, slm}, and the left--right
symmetric models \cite{lrs}.
In the left--right symmetric models the modular 
invariant constraint $\gamma\cdot b_j=0~{\rm mod}~1$ imposes that 
$\gamma\{{\bar\eta}^1;{\bar\eta}^2;{\bar\eta}^3\}=1/2$.
The GSO projection of the basis 
vector $\gamma$ on the states arising from the sectors $b_j$
fixes the vacuum of ${\bar\eta}^j$ with opposite chirality
in the two terms of eq. (\ref{psspinor}). The reason being that the 
combinatorial factor with respect to ${\bar\psi}^{1,\cdots,3}$
is even in the first term and odd in the second, whereas
the $\gamma$ projection that utilises (\ref{su3122breakingbc}) is blind
to ${\bar\psi}^{4,5}$. On the other hand, in the standard--like
\cite{fny, slm} and flipped $SU(5)$ \cite{fsu5} models, that utilise
(\ref{su51breakingbc}), the $\gamma$ projection is not blind to 
${\bar\psi}^{4,5}$, and consequently, in these cases, the vacuum 
of ${\bar\eta}^j$ is fixed with the same chirality for all the states
arising from the sector $b_j$. 

Thus, in models that descend from $SO(10)$
via the $SU(5)\times U(1)$
breaking pattern the charges of a generation from a sector
$b_j$, where $j=1,2,3$, under the corresponding symmetry $U(1)_j$, 
are either $+1/2$ or $-1/2$ for all the states from that sector.
In contrast, in the left--right symmetric string models, the
corresponding charges, up to a sign, are 
\beq
Q_j(Q_L;L_L)=+1/2~~~;Q_j(Q_R;L_R)=-1/2,
\label{u1chargesinlrsmodel}
\eeq
{\it i.e.} the charges of the $SU(2)_L$ doublets have the opposite 
sign from those of the $SU(2)_R$ doublets. 
In fact, this is the reason that, in contrast to the FSU5 \cite{fsu5},
Pati--Salam \cite{alr} and the Standard--like \cite{slm} string models,
in the left--right symmetric models, the $U(1)_j$ symmetries are not
part of the anomalous $U(1)$ symmetry.  
Ultimately, the reason that the
$U(1)_\zeta$ charge assignment in left--right symmetric models is possible
is that, at the level of the NAHE--set, we have spinorial 
16 states with opposite $U(1)_\zeta$ charges. This arises because,
at the level of the NAHE--set, the $SO(10)$ symmetry is not 
enhanced to $E_6$. If the NAHE symmetry is 
extended to $E_6$, the spinorial 16 states with the ``wrong'' $U(1)_\zeta$
charge are projected out. This is also necessarily 
the case in the flipped $SU(5)$ \cite{fsu5} and standard--like models
\cite{slm}, as well as the Pati--Salam models constructed to date \cite{alr}.
By contrast, in the left--right symmetric models,  half of the generation
states are picked from the spinorial 16 with $Q_j=1/2$ and half from 
those with $Q_j=-1/2$. The LRS model given in eqs. (\ref{model1})
and (\ref{phasesmodel1}) is an example of an explicit string model
that exhibit this property. The full massless spectrum of this model,
as well as the superpotential up to quintic order, are given in ref. \cite{lrs}.
 
\LRS Model 1 Boundary Conditions:
\beqn
 &\begin{tabular}{c|c|ccc|c|ccc|c}
 ~ & $\psi^\mu$ & $\chi^{12}$ & $\chi^{34}$ & $\chi^{56}$ &
        $\bar{\psi}^{1,...,5} $ &
        $\bar{\eta}^1 $&
        $\bar{\eta}^2 $&
        $\bar{\eta}^3 $&
        $\bar{\phi}^{1,...,8} $ \\
\hline
\hline
  ${\alpha}$  &  0 & 0&0&0 & 1~1~1~0~0 & 0 & 0 & 0 &1~1~1~1~0~0~0~0 \\
  ${\beta}$   &  0 & 0&0&0 & 1~1~1~0~0 & 0 & 0 & 0 &1~1~0~0~1~1~0~0 \\
  ${\gamma}$  &  0 & 0&0&0 &
${1\over2}$~${1\over2}$~${1\over2}$~0~0&${1\over2}$&${1\over2}$&${1\over2}$ 
&1~${1\over2}$~${1\over2}$~${1\over2}$~${1\over2}$~${1\over2}$~${1\over2}$~0 
\end{tabular}
   \nonumber\\
   ~  &  ~ \nonumber\\
   ~  &  ~ \nonumber\\
     &\begin{tabular}{c|c|c|c}
 ~&   $y^3{y}^6$
      $y^4{\bar y}^4$
      $y^5{\bar y}^5$
      ${\bar y}^3{\bar y}^6$
  &   $y^1{\omega}^5$
      $y^2{\bar y}^2$
      $\omega^6{\bar\omega}^6$
      ${\bar y}^1{\bar\omega}^5$
  &   $\omega^2{\omega}^4$
      $\omega^1{\bar\omega}^1$
      $\omega^3{\bar\omega}^3$
      ${\bar\omega}^2{\bar\omega}^4$ \\
\hline
\hline
$\alpha$& 1 ~~~ 0 ~~~ 0 ~~~ 0  & 0 ~~~ 0 ~~~ 1 ~~~ 1  & 0 ~~~ 0 ~~~ 1 ~~~ 1 \\
$\beta$ & 0 ~~~ 0 ~~~ 1 ~~~ 1  & 1 ~~~ 0 ~~~ 0 ~~~ 0  & 0 ~~~ 1 ~~~ 0 ~~~ 1 \\
$\gamma$& 0 ~~~ 0 ~~~ 1 ~~~ 0  & 1 ~~~ 0 ~~~ 0 ~~~ 1  & 0 ~~~ 1 ~~~ 0 ~~~ 0 \\
\end{tabular}
\label{model1}
\eeqn

\LRS Model 1 Generalized GSO Coefficients:
\begin{equation}
{\bordermatrix{
      &{\bf 1}&  S& & {b_1}&{b_2}&{b_3}& & {\alpha}&{\beta}&{\gamma}\cr
       {\bf 1}&~~1&~~1 & & -1   &  -1 & -1  & & ~~1     & ~~1   & ~~i   \cr
             S&~~1&~~1 & &~~1   & ~~1 &~~1  & &  -1     &  -1   &  -1   \cr
	      &   &    & &      &     &     & &         &       &       \cr
        { b_1}& -1& -1 & & -1   &  -1 & -1  & &  -1     &  -1   & ~~1   \cr
        { b_2}& -1& -1 & & -1   &  -1 & -1  & &  -1     &  -1   & ~~1   \cr
        { b_3}& -1& -1 & & -1   &  -1 & -1  & &  -1     & ~~1   & ~~1   \cr
   	      &   &    & &      &     &     & &         &       &       \cr
      {\alpha}&~~1& -1 & &~~1   &  -1 & -1  & &  -1     & ~~1   & ~~i   \cr
      {\beta }&~~1& -1 & & -1   & ~~1 &~~1  & &  -1     &  -1   & ~~i   \cr
      {\gamma}&~~1& -1 & & -1   &  -1 & -1  & &  -1     & ~~1   & ~~1   \cr}}
\label{phasesmodel1}
\end{equation}

The preservation of the $U(1)$ combination 
\beq
U(1)_\zeta=U_1+U_2+U_3
\eeq
as an anomaly free symmetry is the key to keeping it as an
unbroken proton lifeguard. The left--right symmetric string
models admit cases without any anomalous $U(1)$ symmetry,
and are free of any gauge and gravitational anomalies.
We note that there may exist
string models in the classes of \cite{fsu5, alr, slm} in
which $U(1)_\zeta$ is anomaly free.
This may be the case in the so called self--dual vacua of ref. \cite{fkr}.
In ref. \cite{fkr} a duality symmetry was uncovered in the space
of fermionic $Z_2\times Z_2$ symmetric orbifolds under the exchange of the
total number twisted spinor plus anti--spinor and twisted
vector representations of $SO(10)$. The self--dual models are 
the models in which the total number of 
spinors and anti--spinors is equal to the total number of vector
representations. The self--dual models are free of any $U(1)$ anomalies.
Thus, in such self--dual models with three light chiral 
generations the $U(1)_\zeta$ combination is anomaly free and can
remain unbroken below the string scale. 
Such quasi--realistic self--dual string models, with an anomaly free $U(1)_\zeta$, have
not been constructed to date.

The novel feature of the left--right symmetric string vacua is the 
existence of models in which all the $U(1)$ symmetries are anomaly free.
As discussed above this results from the left--right symmetry breaking pattern
and the assignment of the Standard Model states, which are obtained from the twisted 
sectors $b_j$, and their charges under the $U(1)_j$ symmetries.
In addition to the three light generations arising from the twisted
sectors, the string models contain additional states arising from
the twisted or untwisted sectors. The additional spectrum is
in general highly model dependent. We discuss here, in general terms, this additional
spectrum and the $U(1)_j$ charge assignments in the string models.
Below we will fix our string--inspired model by fitting it with additional states
that are compatible with the string spectrum.  

The twisted sectors $b_j$ can produce additional states that arise
from spinorial representations of the underlying $SO(10)$ symmetry
with charges $\pm1/2$ under $U(1)_j$.  The original $Z_2\times Z_2$ orbifold 
that underlies the free fermionic models has forty--eight fixed points. 
Additional states may therefore arise from different fixed points. 

Sectors that contain the basis vectors that break the $SO(10)$ gauge symmetry,
produce exotic fractionally charged states that must
obtain a sufficiently high mass.
We note the existence of exophobic heterotic string 
models in which fractionally charged states appear only in the massive
string spectrum \cite{exophobic}. We do not consider these states further here. 

The twisted sectors $b_j+x$ produce states that transform in
the vectorial representations of the underlying $SO(10)$ GUT symmetry.  
A twisted sector that produces $SO(10)$ vectorial representations
does not exist in the model of eq. (\ref{model1}). An alternative
model that gives rise to twisted states in the vectorial 
representation of $SO(10)$ is given in eq. (\ref{model2}) 
 
\LRS Model 2 Boundary Conditions:
\beqn
 &\begin{tabular}{c|c|ccc|c|ccc|c}
 ~ & $\psi^\mu$ & $\chi^{12}$ & $\chi^{34}$ & $\chi^{56}$ &
        $\bar{\psi}^{1,...,5} $ &
        $\bar{\eta}^1 $&
        $\bar{\eta}^2 $&
        $\bar{\eta}^3 $&
        $\bar{\phi}^{1,...,8} $ \\
\hline
\hline
  ${\alpha}$  &  0 & 0&0&0 & 1~1~1~0~0 & 0 & 0 & 0 &1~1~1~1~0~0~0~0 \\
  ${\beta}$   &  0 & 0&0&0 & 1~1~1~0~0 & 0 & 0 & 0 &1~1~1~1~0~0~0~0 \\
  ${\gamma}$  &  0 & 0&0&0 &
${1\over2}$~${1\over2}$~${1\over2}$~0~0&${1\over2}$&${1\over2}$&${1\over2}$ 
&0~${1\over2}$~${1\over2}$~${1\over2}$~${1\over2}$~${1\over2}$~${1\over2}$~0 
\end{tabular}
   \nonumber\\
   ~  &  ~ \nonumber\\
   ~  &  ~ \nonumber\\
     &\begin{tabular}{c|c|c|c}
 ~&   $y^3{y}^6$
      $y^4{\bar y}^4$
      $y^5{\bar y}^5$
      ${\bar y}^3{\bar y}^6$
  &   $y^1{\omega}^5$
      $y^2{\bar y}^2$
      $\omega^6{\bar\omega}^6$
      ${\bar y}^1{\bar\omega}^5$
  &   $\omega^2{\omega}^4$
      $\omega^1{\bar\omega}^1$
      $\omega^3{\bar\omega}^3$
      ${\bar\omega}^2{\bar\omega}^4$ \\
\hline
\hline
$\alpha$& 1 ~~~ 0 ~~~ 0 ~~~ 0  & 0 ~~~ 0 ~~~ 1 ~~~ 1  & 0 ~~~ 0 ~~~ 1 ~~~ 1 \\
$\beta$ & 0 ~~~ 0 ~~~ 1 ~~~ 1  & 1 ~~~ 0 ~~~ 0 ~~~ 0  & 0 ~~~ 1 ~~~ 0 ~~~ 1 \\
$\gamma$& 0 ~~~ 0 ~~~ 1 ~~~ 0  & 1 ~~~ 0 ~~~ 0 ~~~ 1  & 0 ~~~ 1 ~~~ 0 ~~~ 0 \\
\end{tabular}
\label{model2}
\eeqn

The sector $b_1+b_2+\alpha+\beta$ in the additive group $\Xi$ spanned 
by this basis gives rise to twisted vectorial $SO(10)$ representations. 
In this sector, the charges under the $U(1)_\zeta$ are fixed
by the vacuum of ${\bar\eta}_1$ and ${\bar\eta}_2$.

In the left--right symmetric models, the twisted sectors $b_j+x$ produce states 
that transform as $SU(2)_L\times SU(2)_R$
bi-doublets with the $U(1)_\zeta$ charge assignments $$(1,2,2,0,\pm1),$$
as well as colour triplets. 
The $U(1)_\zeta$ charges of these colour triplets  depend
on the $\gamma$ projection and there are several possibilities.
If the twisted plane produces bi--doublets with $+1$ $U(1)_\zeta$ charge,
then the $\gamma$ projection dictates that any colour triplet arising
from that sector is neutral under $U(1)_\zeta$. In this case, 
we must take the colour triplets to have the charges 
$$(3,1,1,-1,0) ~+~({\bar3},1,1,+1,0).$$
The vectorial states arising from the twisted sector depend, however,
on the specific choice of basis vectors, and the pairing of the 
world--sheet fermions from the set $\{y, \omega|{\bar y},
{\bar\omega}\}$ into complex, or Ising, type fermions, as well 
as on the GGSO projection coefficients. There exist choices 
of basis sets that produce vectorial states from none, one, two or
three of the twisted planes. To date, only models of the 
first and second class have been studied in detail, where
the example in eq. (\ref{model1}) belongs to the first kind,
and the example in eq. (\ref{model2}) belongs to the second.
If the twisted plane produces both electroweak doublets and color
triplets, the $\gamma$--projection dictates that they have 
$\pm1$ and vanishing $U(1)_\zeta$ charges, respectively, 
and vice versa. 

An alternative possibility, is that more than one twisted plane
produces states in vectorial $SO(10)$ representations. In this 
case, one plane can produce bi--doublets and a second plane produces
the colour triplets. Here, the charges of the twisted colour 
triplets are not correlated with those of the bi--doublets 
and we can obtain twisted vectorial colour triplets with charges 
$$(3,1,1,+1,-1) ~+~({\bar3},1,1,-1,+1)$$ 
or 
$$(3,1,1,+1,+1) ~+~({\bar3},1,1,-1,-1).$$ 

Electroweak Higgs bi--doublets may also arise from the untwisted sector. 
However, in this case the $\gamma$ GGSO projection
dictates that the untwisted Higgs bi--doublets are neutral under 
$U(1)_\zeta$, 
\beq
h       = (1,2,2,0,0) ~~=~  
{\left(\matrix{
                 h^u_+  &  h^d_0\cr
                 h^u_0  &  h^d_-\cr}\right)}~~. 
\label{HiggsLRsymreps}
\eeq
The untwisted Higgs bi--doublet is the one that 
forms invariant leading mass terms with the Standard Model matter states,
due to the fact that the $Q_L$ and $Q_R$ multiplets carry opposite
$U(1)_\zeta$ charge.

The string models may also produce $SO(10)$ singlets, which carry
$U(1)_\zeta$ charges that are compatible with
the string charge assignments. The singlets can arise 
from the Neveu--Schwarz untwisted sector and twisted
sectors that produce vectorial representations, like 
the sector $b_1+b_2+\alpha+\beta$ in the model
generated by eq. (\ref{model2}). The $U(1)_\zeta$ charges 
are fixed according to the following rules. In the untwisted sector
these states arise by acting on the vacuum with two oscillators
$\bar\eta^i$ and $\bar\eta^j$. Their $U(1)_\zeta$
charges are fixed by the $\gamma$ projection 
according to the sign of $\delta_\gamma$ in eq. (\ref{gsoprojections}), 
being zero for $\delta_\gamma=+1$ and $\pm2$ for $\delta_\gamma=-1$.
In the twisted sector we consider for concreteness 
the sector $b_1+b_2+\alpha+\beta$ in the model spanned by
eq. (\ref{model2}). The singlets from that sector are obtained
by acting on the vacuum with $\bar\eta^3$, or with an oscillator 
of a real fermion from the set $\{ {\bar y}{\bar\omega}\}$,
which are not periodic in $b_1+b_2+\alpha+\beta$.
The $U(1)_\zeta$ charges are again fixed by the $\gamma$ projection.  
The $\gamma$ GGSO projection phase in this sector
can be either $\pm1$ or $\pm i$. Depending on this GGSO phase and
the type of state, the $U(1)_\zeta$ charges in this sector
can be $\pm2$, $\pm1$ or zero. Therefore, we can have a combination 
of singlets with charges $+2$ and $+1$, 
as well as singlets with vanishing $U(1)_\zeta$ charge.

\section{Anomaly analysis and string--inspired models}

The $U(1)_\zeta$ symmetry forbids the dimension four, five and six proton 
decay mediating operators \cite{ps}. It arises as an anomaly 
free symmetry in the string models. We need to ensure that 
it remains anomaly free in the low energy effective 
field theory. For this purpose we construct a string--inspired 
model that takes into account the $U(1)$ charges of the Standard Model
matter states as they arise in the string model, and we augment 
the model with additional states, compatible with the 
string charge assignments, to render the spectrum of
the string--inspired effective field theory anomaly free. 
In terms of the left--right symmetric decomposition of ref. \cite{lrs},
the embedding of the Standard Model matter states 
is in the following representations:
\beqn
Q_L &=& (      3 ,2,1,~{+{1\over2}},  {-{1\over2}})~~,  
\label{QL}\\
Q_R &=& ({\bar 3},1,2, {-{1\over2}},  {+{1\over2}}) ~=~  {U+D}~~,
\label{QR}\\
L_L &=& (      1 ,2,1, {-{3\over2}},  {-{1\over2}}) ~~,
\label{LL}\\
L_R &=& (      1 ,1,2,~{+{3\over2}},  {+{1\over2}}) ~~=~  {E+N}~~,
\label{LR}
\eeqn
of $SU(3)\times SU(2)_L\times SU(2)_R\times U(1)_C\times U(1)_\zeta$.
These states arise from the twisted sectors in the string models. The
light Standard Model Higgs representations arise from the untwisted 
sector and transform as $(1,2,2,0,0)$ under this group. 
To construct a consistent low scale model we need to consider the following 
anomalies: 

\begin{description}
\item [${{\cal A}_1:\left(SU(3)_C^2\times U(1)_\zeta\right)}$ ] Only quarks are summed over in this diagram and we find that for the Standard Model fields it is anomaly free.

\item[${{\cal A}_2:\left(SU(2)_L^2\times U(1)_\zeta\right)}$] Due to our charge assignment, the left-handed quark and lepton fields have the same sign resulting in an anomaly.  In fact, 
${\cal A}_2^{SM}=-2$.

\item[${{\cal A}_3:\left(SU(2)_R^2\times U(1)_\zeta\right)}$] Again, because of our charge assignment for right-handed quark and lepton fields there is a resulting anomaly.  In fact, 
${\cal A}_3^{SM}=+2$.

\item[${{\cal A}_4:\left(U(1)_C^2\times U(1)_\zeta\right)}$] All fermions are summed over in this diagram.  As left- and right-handed fields have opposite charge it is found to be anomaly free.  

\item[${{\cal A}_5:\left(U(1)_C\times U(1)_\zeta^2\right)}$] Again all fermions are summed over in this diagram.  It is also found to be anomaly free due to the opposite charge assignments for left- and right-handed fields.  

\item[${{\cal A}_6:\left(U(1)_\zeta\times \mbox{Gravity}\right)}$] Here we also sum over all fermions. Due to our choices of $Q_\zeta$ this diagram is obviously anomaly free. 

\item[${{\cal A}_7:\left(U(1)_\zeta^3\right)}$] Again we sum over all fermions and the diagram is found to be anomaly free.
\end{description}

We note that with the charge assignment in eqs. (\ref{QL}--\ref{LR}), 
the spectrum possesses mixed $SU(2)_{L,R}^2\times U(1)_\zeta$ anomalies.
In the string vacua, these anomalies are canceled
by additional states that arise in the string models. 
The string vacua are therefore entirely free of
gauge and gravitational anomalies. However, the 
additional spectrum in the string vacua is highly model
dependent. We therefore judicially augment the spectrum
in eqs. (\ref{QL}--\ref{QR}) with additional states
that cancel the $SU(2)_{L,R}^2\times U(1)_\zeta$ mixed 
anomalies. This guarantees that any combination
of the $U(1)$ generators in the Cartan subalgebra 
of the $SU(3)\times SU(2)_L\times SU(2)_R\times U(1)_C\times U(1)_\zeta$
gauge group is anomaly free. That is it guarantees that any
$Z^\prime$ arising from this group is anomaly free at the low scale. 
To obtain a spectrum which is free of the mixed anomalies we
add to each generation two copies of the states with charges
\beqn
H_L        &=& (      1 ,2,1,~{+{3\over2}},  {+{1\over2}})~~,  
\label{HL}\\
H_L^\prime &=& (      1 ,2,1, {-{3\over2}},  {+{1\over2}}) ~~,
\label{HLp}\\
H_R        &=& (      1 ,1,2, {+{3\over2}},  {-{1\over2}}) ~~,
\label{HR}\\
H_R^\prime &=& (      1 ,1,2,~{-{3\over2}},  {+{1\over2}}) ~~,
\label{HRp}
\eeqn
With this augmentation the spectrum is free of all gauge and gravitational anomalies. 

In addition to the light spectrum,
heavy Higgs states in vector--like representations are needed to
break the $$SU(2)_R\times U(1)_C\times U(1)_\zeta\rightarrow
U(1)_Y\times U(1)_{Z^\prime}.$$ These are
$$
{\mathcal{H}}_{R}+\bar{\mathcal{H}}_{R}=(      1 ,1,2,~{+{3\over2}},  {-{1\over2}})
+ (1 ,1,2,~{-{3\over2}},  {+{1\over2}}).
$$
The weak hypercharge is given by the combination
$$U(1)_Y=T_{3_R}+{1\over3}U(1)_C~,$$
where $T_{3_R}$ is the diagonal generator of $SU(2)_R$,
and the electromagnetic $U(1)$ current given by the combination
$$U(1)_{\it e.m.}= T_{3_L}+U_Y.$$
The VEV of the neutral component in $\langle {\mathcal{H}}_{R}\rangle =
\langle \bar{\mathcal{H}}_{R}\rangle$ leaves the unbroken $U(1)_{Z^\prime}$
combination given by
\beq
U(1)_{Z^\prime}={1\over5}U_C-{2\over5}T_{3_R}+U_\zeta.
\label{u1zprimeplus}
\eeq
The augmentation of the states in eqs (\ref{QL}--\ref{LR}) with the states 
given by eqs. (\ref{HL}--\ref{HRp}) guarantees that the effective 
low energy field theory below the intermediate breaking scale 
is completely free of gauge and gravitational anomalies. 
Alternatively, we can take the heavy Higgs fields as 
$${\mathcal{H}}_{R}+\bar{\mathcal{H}}_{R}=(      1 ,1,2,~{+{3\over2}},  {+{1\over2}})
+ (1 ,1,2,~{-{3\over2}},  {-{1\over2}}).$$
This choice leaves the unbroken $U(1)_{Z^\prime}$
combination given by
\beq
U(1)_{Z^\prime}={1\over5}U_C-{2\over5}T_{3_R}-U_\zeta.
\label{u1zprimeminus}
\eeq
For definiteness we will focus in this paper on the choice given in 
eq. (\ref{u1zprimeplus}). 
In addition to the electroweak doublets that are chiral with respect to
$U(1)_\zeta$ and that are needed to cancel the 
$SU(2)_{L,R}^2\times U(1)_\zeta$ anomalies, the 
models may contain additional colour triplets and $SO(10)$ singlets in 
vector--like representations. The colour triplets may be needed to facilitate 
compatibility of the heterotic string coupling unification with the low energy
gauge sector data. We defer a detailed analysis of this issue to future work
and include here the additional triplets for completeness. $SO(10)$ singlets 
that are charged under $U(1)_\zeta$, are required in the low energy field 
theory to break the $U(1)_{Z^\prime}$ gauge symmetry.


%

\begin{table}[!h]
\noindent 
{\small
\openup\jot
\begin{center}
\begin{tabular}{|l|ccc|c|c|}
\hline
Field&$\hphantom{\times}SU(3)_C$&$\times SU(2)_L $&$\times SU(2)_R$&${U(1)}_C$&${U(1)}_\zeta$\\[1ex]
\hline
$Q_L^i$&$3$&$2$&$1$&$+\frac{1}{2}$&$-\frac{1}{2}$\\[1ex]
$Q_R^i$&$\bar{3}$&$ 1$&$ 2$&$-\frac{1}{2}$&$+\frac{1}{2}$\\[1ex]
$L_L^i$&$1$&$2$&$1$&$-\frac{3}{2}$&$-\frac{1}{2}$\\[1ex]
$L_R^i$&$1$& $1$&$ 2$&$+\frac{3}{2}$&$+\frac{1}{2}$\\[1ex]
\hline
$H_0$&$1$&$ 2$& $2$&$\hphantom{+}0$&$\hphantom{+}0$\\[1ex]
\hline
$H_{L}^{ij}$&$1$&$2$&$1$&$+\frac{3}{2}$&$+\frac{1}{2}$\\[1ex]
$H_{L}^{\prime \hphantom{\prime} ij}$&$1$&$2$&$1$&$-\frac{3}{2}$&$+\frac{1}{2}$\\[1ex]
$H_{R}^{ij}$&$1$&$1$&$2$&$-\frac{3}{2}$&$-\frac{1}{2}$\\[1ex]
$H_{R}^{\prime \hphantom{\prime}ij}$&$1$&$1$&$2$&$+\frac{3}{2}$&$-\frac{1}{2}$\\[1ex]
\hline
$D^n$&$3$&$1$&$1$&$+1$&$\hphantom{+}0$\\[1ex]
$\bar{D}^n$&$\bar{3}$&$1$&$1$&$-1$&$\hphantom{+}0$\\[1ex]
\hline
${\cal H}_R$&$1$&$1$&$2$&$+\frac{3}{2}$&$-\frac{1}{2}$\\[1ex]
$\bar{\cal H}_R$&$1$&$1$&$2$&$-\frac{3}{2}$&$+\frac{1}{2}$\\[1ex]
\hline

$S^i$&$1$&$1$&$1$&$\hphantom{+}0$&$-1$\\[1ex]
$\bar{S}^i$&$1$&$1$&$1$&$\hphantom{+}0$&$+1$\\[1ex]
$\phi^a$&$1$&$1$&$1$&$\hphantom{+}0$&$\hphantom{-}0$\\[1ex]
\hline
\end{tabular}\end{center}
}
\caption{\label{table1}\it
High scale spectrum and
$SU(3)_C\times SU(2)_L\times SU(2)_R\times U(1)_C\times U(1)_E$ 
quantum numbers, with $i=1,2,3$ for the three light 
generations, $j=1,2$ for the number of doublets required by 
anomaly cancellation, $n=1,...,k$, and $a=1,...,p$.}
\end{table}

The spectrum of our model above the left--right symmetry breaking scale is 
summarised in table \ref{table1}. The spectrum below the intermediate symmetry 
breaking scale is shown in table \ref{table3}. The spectra above and below the 
symmetry breaking scale are both free all gauge and gravitational anomalies. 
Hence, the $U(1)_{Z^\prime}$ combination given in eq.
(\ref{u1zprimeplus}) is viable at 
low energies. It is family universal and hence is not constrained by
flavour changing neutral currents. The superpotential above the intermediate 
symmetry breaking scale is shown in eq. (\ref{leftrightsymmetricsup}),
\beqn
W = 
   &&\lambda_{1}Q_LQ_RH_0 
     + \lambda_{2} L_LL_RH_0 
     + \lambda_{3}H_{L}H_{R}H_0\nonumber\\ 
+ &&\lambda_{4}H_{L}^{\prime}H_{R}^{\prime}H_0 
    + \lambda_{5} D\bar{D}\phi 
    +\lambda_{6}  S\bar{S}\phi 
    +\lambda_{7}H_{L}H_{L}^{\prime}S \nonumber\\
+ &&\lambda_{8}H_{R}H_{R}^{\prime}\bar{S}
   + \lambda_{9}\phi\phi\phi
   +\mu H_0H_0\label{leftrightsymmetricsup}\\ 
+ &&  \eta_{1}L_R\bar{\cal H}_RS
        + \eta_{2}H_{R}{\cal H}_R\bar{S}
        + \eta_{3}H_{L}^{\prime}{\cal H}_RH_0
        + \eta_{4}H_{R}^{\prime}\bar{\cal H}_R\phi 
        + \eta_{5} {\cal H}_R\bar{\cal H}_R\phi~,\nonumber
\eeqn
where indices have been suppressed and the couplings labeled by $\eta_j$ are those that involve the 
couplings to the Higgs fields that break the left--right symmetry. 
The first two terms produce Dirac masses for the quark and leptons. 
A Dirac mass term for the neutrino is admitted due to the 
left--right symmetry. The model admits a type III seesaw mechanism 
by the couplings in $\eta_1$ and $\lambda_9$. 
We note that the intermediate scale breaking is a free parameter in this model
as it is not constrained by the doublet--triplet
splitting, which is induced at the string level \cite{pstudies}. Hence,
the only constraint are imposed by the masses
of the left--handed neutrinos. These can be sufficiently suppressed by the 
type III seesaw mechanism, and by rendering the left--handed neutrinos 
unstable by the coupling to light sterile neutrinos $\phi_m$. 
A detailed analysis of the neutrino mass spectrum and the constraints on 
the intermediate symmetry breaking scale is deferred to future work. 
We included in the spectrum colour triplet fields, in vector--like 
representations, which may be needed 
to facilitate gauge coupling unification at the string scale.
Non--Abelian singlet fields in vector like representations that
carry $U(1)_{Z^\prime}$ charge are required to break the $U(1)_{Z^\prime}$
gauge symmetry. Finally, singlets of the entire low scale gauge 
symmetry are included. Such states may arise, for example, from hidden 
sector condensates in the string models. The renormalization group evolution 
of the gauge and superpotential couplings, together with hidden sector 
dynamics are expected to fix all scales in the string models. 

Returning to the superpotential eq. (\ref{leftrightsymmetricsup}), the 
couplings $\lambda_3$--$\lambda_9$ involve couplings of the extra 
doublets, triplets and singlets in the model, and the $\mu$ parameter
is the usual supersymmetric Higgs parameter. The couplings in the 
last row in eq. (\ref{leftrightsymmetricsup}) are those that involve 
the couplings to the heavy Higgs fields. We note that the 
choice given in eq. (\ref{u1zprimeplus}) forbids the Higgsino--neutrino
mixing term $L_L{\bar {\cal H}}H_0$ at the expense that the $\nu_L^c$ fields
are charged under $U(1)_{Z^\prime}$, whereas the choice given 
in eq. (\ref{u1zprimeminus}) allows the neutrino--Higgsino
mixing term, but keeps the $\nu_L^c$ fields neutral under 
the $U(1)_{Z^\prime}$ combination. This issue again relates
to the scale of $SU(2)_R$ breaking and the consequent
suppression of the left--handed neutrino masses. We will examine
this question in more detail in future work. We note here that some 
couplings in eq. (\ref{leftrightsymmetricsup})
may still need to be suppressed to avoid conflict with the data. 

Turning to the proton decay mediating operators we note that with
both choices in (\ref{u1zprimeplus}) and (\ref{u1zprimeminus})
the dimension four baryon number violating operator
that arise from 
\beq
{Q_RQ_RQ_R{\cal H}_R} ~~\rightarrow~~   \{UDD{\cal N}\} \label{uddH}
\eeq
as well as the 
dimension five baryon number violating operator
\beqn
{Q_LQ_LQ_LL_L} & \rightarrow & QQQL \label{qqql}\\
{Q_RQ_RQ_RL_R} & \rightarrow &  \{UDDN,UUDE\} \label{uddn}
\eeqn
are forbidden by $U(1)_{Z^\prime}$.
The lepton number violating operators that arise from
\beqn
{Q_LQ_RL_L{\cal H}_R} & \rightarrow & QDL\cal{N} \label{qdln}\\
{L_LL_LL_R{\cal H}_R} & \rightarrow & LLE\cal{N} \label{llen}
\eeqn
are also forbidden for the model of eq. (\ref{u1zprimeplus}). 
For the model of eq. (\ref{u1zprimeminus}), the lepton number 
violating operators are allowed.  Hence, the proton decay mediating 
operators are suppressed by $\Lambda_{Z^\prime}/M_{\rm Planck}$, which
yields adequate suppression provided that the $U(1)_{Z^\prime}$ breaking
scale is sufficiently low as discussed in \cite{ps}. 

\begin{table}[!h]
\noindent 
{\small
\openup\jot
\begin{center}
\begin{tabular}{|l|cc|c|c|c|}
\hline
Field&$\hphantom{\times}SU(3)_C$&$\times SU(2)_L $&$T_{3R}$&${U(1)}_Y$&${U(1)}_{Z^{\prime}}$\\[1ex]
\hline
$Q_L^i$&$3$&$2$&$\hphantom{+}0$&$+\frac{1}{6}$&$-\frac{2}{5}$\\[1ex]
$u_L^{c\hphantom{\prime}i}$&$\bar{3}$&$1$&$-\frac{1}{2}$&$-\frac{2}{3}$&$+\frac{3}{5}$\\[1ex]
$d_L^{c\hphantom{\prime}i}$&$\bar{3}$&$1$&$+\frac{1}{2}$&$+\frac{1}{3}$&$+\frac{1}{5}$\\[1ex]
$L_L^i$&$1$&$2$&$\hphantom{+}0$&$-\frac{1}{2}$&$-\frac{4}{5}$\\[1ex]
$e_L^{c\hphantom{\prime}i}$&$1$&$1$&$-\frac{1}{2}$&$+1$&$+\frac{3}{5}$\\[1ex]
$\nu_L^{c\hphantom{\prime}i}$&$1$&$1$&$+\frac{1}{2}$&$\hphantom{+}0$&$+1$\\[1ex]
\hline
$H^u$&$1$&$2$&$+\frac{1}{2}$&$+\frac{1}{2}$&$-\frac{1}{5}$\\[1ex]
$H^d$&$1$&$2$&$-\frac{1}{2}$&$-\frac{1}{2}$&$+\frac{1}{5}$\\[1ex]
\hline
$H^i_{L}$&$1$&$2$&$\hphantom{+}0$&$+\frac{1}{2}$&$+\frac{4}{5}$\\[1ex]
$H^{\prime\hphantom{\prime}i}_{L}$&$1$&$2$&$\hphantom{+}0$&$-\frac{1}{2}$&$+\frac{1}{5}$\\[1ex]
$E^{i}_{R}$&$1$&$1$&$-\frac{1}{2}$&$-1$&$-\frac{3}{5}$\\[1ex]
$N^i_{R}$&$1$&$1$&$+\frac{1}{2}$&$\hphantom{+}0$&$-1$\\[1ex]
$E{\prime\hphantom{\prime}i}_{R}$&$1$&$1$&$+\frac{1}{2}$&$+1$&$-\frac{2}{5}$\\[1ex]
$N^{\prime\hphantom{\prime}i}_{R}$&$1$&$1$&$-\frac{1}{2}$&$\hphantom{+}0$&$\hphantom{+}0$\\[1ex]
\hline
$D^n$&$3$&$1$&$\hphantom{+}0$&$+\frac{1}{3}$&$+\frac{1}{5}$\\[1ex]
$\bar{D}^n$&$\bar{3}$&$1$&$\hphantom{+}0$&$-\frac{1}{3}$&$-\frac{1}{5}$\\[1ex]
\hline

$S^i$&$1$&$1$&$\hphantom{+}0$&$\hphantom{+}0$&$-1$\\[1ex]
$\bar{S}^i$&$1$&$1$&$\hphantom{+}0$&$\hphantom{+}0$&$+1$\\[1ex]
$\phi^a$&$1$&$1$&$\hphantom{+}0$&$\hphantom{-}0$&$\hphantom{-}0$\\[1ex]
\hline
\end{tabular}
\end{center}
}
\caption{\label{table3}\it
Low scale matter spectrum and
$SU(3)_C\times SU(2)_L\times U(1)_Y\times U(1)_{Z^{\prime}}$ quantum numbers. }
\end{table}

In table \ref{table3} and eq. (\ref{lowscalesup}) we display the superpotential 
below the intermediate symmetry breaking scale. The model offers novel experimental 
signatures at contemporary colliders that will be studied in forthcoming publications. 

\beqn
W_0= &&h_{u}Q_Lu_L^{c}H^u 
               +h_{d}Q_Ld^{c}_LH^d
              + h_{e}L_Le_L^{c}H^d
              + h_{\nu}L_L\nu_{L}^{c}H^u \nonumber \\ \nonumber
          &&+\lambda_{H}H_{L}N_{R}H^d
              + \lambda^{\prime}_{H}H_{L}E_{R}H^u\\ \nonumber
         + &&\lambda_{H^{\prime}}H_{L}^{\prime}E_{R}^{\prime}H^d 
              + \lambda^{\prime}_{H^{\prime}}H^{\prime}_{L}N_{R}^{\prime}H^u 
              + \lambda_{1}H_{L}H_{L}^{\prime}S \\
        + &&\lambda_{2}E_{R}E_{R}^{\prime}\bar{S} 
             +\lambda^{\prime}_{2}N_{R}N_{R}^{\prime}\bar{S}
             +\mu H^uH^d\label{lowscalesup}
\eeqn
where all indices have been suppressed.

\section{Conclusions}

The structure of the Standard Model spectrum and gauge charges motivates the
embedding of its matter states into representations of a GUT group, 
in particular 
into those of $SO(10)$. Proton lifetime constraints, on the other hand, 
indicate that the unification scale must be vastly separated from the 
electroweak scale, to adequately suppress the proton decay mediating operators.
Augmenting the GUT theory with supersymmetry then provides the means 
to connect the two vastly separated scales in a perturbatively controlled 
framework. This augmentation, however, introduces new dimension four 
and five proton decay mediating operators. It is important to emphasize
that proton decay mediating operators are expected to arise in most 
extensions of the Standard Model. The reason being that in
the Standard Model baryon and lepton numbers
are accidental global symmetries at the renormalizable level.
Extensions of the Standard Model introduce an effective cutoff scale.
We expect that all operators which are compatible with the Standard Model
gauge symmetries are generated, unless they are forbidden by a local 
or a discrete local symmetry. An appealing proposition is that the proton
decay mediating operators are forbidden by a new Abelian gauge symmetry.
To provide adequate suppression, this Abelian symmetry has to be broken
at an intermediate scale \cite{ps}, 
possibly within reach of contemporary experiments. 

String theory provides a unique framework for the unification of gravity 
and the gauge interactions, while heterotic string theory further admits the 
$SO(10)$ unification structures that are motivated by the Standard Model data.
Three generation heterotic string derived models, that admit the $SO(10)$
embedding of the Standard Model spectrum, have been studied in the 
free fermionic formulation in the past two decades.

Abelian extensions of the Standard Model have been amply discussed 
in string--inspired scenarios. It is instructive to explore what
lessons might be gleaned from phenomenological free fermionic string models.
The main lesson to be learned is that most of the extra $U(1)$ symmetries
that arise in string models must be broken near the string scale. 
The reason being that these models typically contain an anomalous
$U(1)$ that breaks supersymmetry near the Planck scale. Restoration
of supersymmetry typically implies that all the $U(1)$ symmetries are 
broken. 

Models that may yield an unbroken $U(1)$ symmetry are 
those in which all the extra $U(1)$s are anomaly free. 
Three generation free fermionic models with this property
are those in which the $SO(10)$ symmetry is broken to 
the left--right symmetric subgroup. These models are therefore
supersymmetric and completely free of gauge and gravitational
anomalies. The $U(1)_\zeta$ symmetry in the string models
is an anomaly free, family universal symmetry that forbids
the dimension four, five and six proton decay mediating 
operators, while allowing for the Standard Model fermion
mass terms. A unbroken combination of $U(1)_\zeta$ together with 
$U(1)_{B-L}$ and $U(1)_{T_{3_R}}$ remains unbroken down to 
low energies. It forbids baryon number violation while allowing
for lepton  number violation. Hence, it allows for generation 
of small left--handed neutrino masses via a see saw mechanism. 
Proton decay mediating operators are only generated when the
$U(1)_{Z^\prime}$ is broken. Hence, the scale of the $U(1)_{Z^\prime}$
breaking is constrained by proton lifetime limits and can be
within reach of the contemporary experiments.

Considering only the Standard Model states, $U(1)_\zeta$ 
has mixed $SU(2)_{L,R}$ anomalies. These anomalies are compensated by additional states that
arise in the string models. However, this additional spectrum
is highly model dependent. Our aim in this paper was to construct
string--inspired models that: incorporate the additional 
$U(1)_\zeta$ symmetry; include additional states that 
are compatible with the string charge assignments; 
and are free of gauge and gravitational anomalies. 
We presented two such models, which differ by the particular
unbroken $U(1)_{Z^\prime}$ combination. The models suggest
various phenomenological implications that will be
studied in future publications. 

The main property of $U(1)_\zeta$ in the left--right symmetric
free fermionic models is that it is anomaly free. 
Among the quasi--realistic free fermionic models 
only the left--right symmetric models produce a
models in which all the $U(1)$ symmetries are anomaly free.
In other free fermionic models the symmetry breaking pattern $E_6\rightarrow 
SO(10)\times U(1)_E$ results with an anomalous $U(1)_E$. Extra 
$U(1)$ symmetries inspired from $E_6$ have been amply discussed
in the literature \cite{so10e6zprime}. The question then arises can 
an anomaly free $U(1)_E$ arise in the string models? In general
in free fermionic models, the answer would be negative. A
possible exception can be the case of the self--dual models
under the spinor--vector duality of ref. \cite{fkr}. In the
self--dual models the three spinorial 16 representations
that are used to accommodate the Standard Model states
are accompanied by three vectorial 10 representations and 
the corresponding singlets. Thus, these models retain the
$E_6$ embedding of the spectrum, but project the space--time
vector states that enhance $SO(10)$ to $E_6$. This is possible
if the spinorial and vectorial states are obtained from
different fixed points. Thus, while the spectrum possesses 
an $E_6$ embedding and $U(1)_E$ is anomaly free,
the gauge symmetry is $SO(10)$ and is not enhanced to $E_6$.

\section{Acknowledgments}

AEF would to thank the University of Oxford for hospitality. 
This work was supported in part by the STFC (PP/D000416/1).


\begin{thebibliography}{99}

\bibitem{psreview} For review and references see {\it e.g.}:\\
                      P.~Nath, P.~Fileviez Perez, \PRT{441}{2007}{191}.

\bibitem{qge} S. Hawking, D. Page and C. Pope, 
			 \PLB{86}{1979}{175}; \NPB{170}{1980}{283};\\
		J. Ellis, J.S. Hagelin, D.V. Nanopoulos and K.A.
                                Tamvakis, \PLB{124}{1983}{484};\\
		S. Giddings and A. Strominger, \NPB{307}{1988}{854};\\
                G. Gilbert, \NPB{328}{1989}{159};\\
		F.C. Adams, G.L. Kane, M. Mbonye and M.J. Perry,
				\IJMP{16}{2001}{2399}.

\bibitem{ps} \AEF, \PLB{499}{2001}{147};\\
             C.~Coriano, \AEF~ and M.~Guzzi, \EJP{C53}{2008}{421}.

\bibitem{wei} S. Weinberg, \PRD{26}{1982}{287};\\
 	      N. Sakai and T. Yanagida, \NPB{197}{1982}{533}.

\bibitem{patisalam} J.C. Pati and A. Salam, \PRD{10}{1974}{275}.

\bibitem{dmr} K.R. Dienes and J. March--Russell, \NPB{479}{1996}{113}.

\bibitem{custodial} \AEF, \PLB{339}{1994}{223}.
 
\bibitem{pati} J. Pati, \PLB{388}{1996}{532}.

\bibitem{zprimeliterature} 
  J.L.~Hewett, T.G.~Rizzo, \PRT{183}{1989}{193};\\
  A.~Leike, \PRT{317}{1999}{143};\\
  P.~Langacker, \RMP{81}{2009}{1199}.

\bibitem{Zwirner}F. Zwirner, \IJMP{A3}{1988}{49}.

\bibitem{so10e6zprime}
  G.~Costa, J.R.~Ellis, G.L.~Fogli, D.V.~Nanopoulos, F.~Zwirner,
  \NPB{297}{1988}{244};\\
  S.F.~King, S.~Moretti, R.~Nevzorov, \PRD{73}{2006}{035009};\\
  P.~Athron, S.F.~King, D.J.~Miller, S.~Moretti, R.~Nevzorov, \PRD{80}{2009}{035009}.

\bibitem{fsu5} I.\ Antoniadis, J.\ Ellis, J.\ Hagelin and D.V.\ Nanopoulos
                \PLB{231}{1989}{65}.

\bibitem{fny} \AEF, D.V.\ Nanopoulos and K.\ Yuan,
                                                 \NPB{335}{1990}{347};\\
	      \AEF, \PRD{46}{1992}{3204}.

\bibitem{alr} I.\ Antoniadis, G.K.\ Leontaris and J.\ Rizos,
				\PLB{245}{1990}{161};\\
		G.K.\ Leontaris and J.\ Rizos,
				\NPB{554}{1999}{3}. 

\bibitem{slm}A.E.\ Faraggi, \PLB{278}{1992}{131};
			   \NPB{387}{1992}{239}; \\
        	G.B.\ Cleaver, \AEF~ and D.V.\ Nanopoulos,
                       \PLB{455}{1999}{135};\\
		\AEF, E. Manno and C. Timirgaziu, \EJP{C50}{2007}{701}.

\bibitem{lrs} G.B.\ Cleaver, A.E.\ Faraggi and C.\ Savage,
                                \PRD{63}{2001}{066001};\\
              G.B.\ Cleaver, D.J.\ Clements and A.E.\ Faraggi,
                        \PRD{65}{2002}{106003}.	

\bibitem{exophobic} B. Assel \etal, \PLB{683}{2010}{306}; 
                                    \NPB{844}{2011}{365};\\
        K. Christodoulides, \AEF~ and J. Rizos, arXiv:1104.2264 [hep-ph].

\bibitem{xmap} \AEF, \NPB{407}{1993}{57}; \PLB{326}{1994}{62}.

\bibitem{lepzprime} \AEF~ and D.V. Nanopoulos, \MODA{6}{1991}{61}.

\bibitem{cleaverau1} G.B. Cleaver and \AEF, \IJMP{14}{1999}{2335}.

\bibitem{fff}  H.\ Kawai, D.C.\ Lewellen, and S.H.-H.\ Tye,
					\NPB{288}{1987}{1};\\
               I.\ Antoniadis, C.\ Bachas, and C.\ Kounnas,
	       \NPB{289}{1987}{87};\\
	       I.\ Antoniadis and C.\ Bachas, \NPB{289}{1987}{87}.

\bibitem{nahe} \AEF~and D.V.\ Nanopoulos, \PRD{48}{1993}{3288};\\
               \AEF, \IJMP{14}{1999}{1663}.

\bibitem{fkr} \AEF, C. Kounnas, S. Nooij and J. Rizos, \NPB{695}{2004}{41};\\
		\AEF, C. Kounnas and J. Rizos, \PLB{648}{2007}{84};
					       \NPB{774}{2007}{208};
                                               \NPB{799}{2008}{19};\\
              T. Catelin-Jullien, \AEF, C. Kounnas and John Rizos,
              \NPB{812}{2009}{103}.

\bibitem{su421}G.B.\ Cleaver, A.E.\ Faraggi and S.E.M.\ Nooij, 
		\NPB{672}{2003}{64}.

\bibitem{pstudies} \AEF, \NPB{428}{1994}{111}; 
			\PLB{520}{2001}{337}.


\end{thebibliography}
\end{document}